\documentclass[aps,prd,showpacs,preprintnumbers,onecolumn]{revtex4}
\usepackage{epsfig}
\usepackage{latexsym,amsmath,amssymb,amstext}

\begin{document}
\title{Anomalies at finite density and chiral fermions}
\author{R.\ V.\ \surname{Gavai}}
\email{gavai@tifr.res.in}
\affiliation{Department of Theoretical Physics, Tata Institute of Fundamental
         Research,\\ Homi Bhabha Road, Mumbai 400005, India.}
\author{Sayantan\ \surname{Sharma}}
\email{ssharma@theory.tifr.res.in}
\affiliation{Department of Theoretical Physics, Tata Institute of Fundamental
         Research,\\ Homi Bhabha Road, Mumbai 400005, India.}

\begin{abstract}
Using perturbation theory in the Euclidean (imaginary time) formalism as
well as the nonperturbative Fujikawa method, we verify that the chiral
anomaly equation remains unaffected in the presence of nonzero chemical
potential, $\mu$.  We extend our considerations to fermions with exact chiral 
symmetry on the lattice and discuss the consequences for the recent 
Bloch-Wettig proposal for the Dirac operator at finite chemical potential. 
We propose a new simpler method of incorporating $\mu$ and compare it with
the Bloch-Wettig idea.
\end{abstract}
\preprint{TIFR/TH/09-18}
\maketitle
\section{Introduction}
As we know from Noether's theorem, invariance of a Lagrangian of a
classical field theory under a continuous symmetry leads to conserved
currents.   Inclusion of quantum loop corrections can, however, make some
currents anomalous, and thus lead to the breaking of the
corresponding symmetry.  Chiral anomalies are a well-known example of this
phenomenon.  Chiral anomalies arise in a theory of massless fermions
interacting with the gauge fields . The flavorless axial current of the
fermions is classically conserved but is violated at one-loop level, as was
shown in the famous calculation of the Adler-Bell-Jackiw(ABJ) triangle
diagram for the $U(1)$ case \cite{adler,bell}. The anomalous contribution
is a universal feature of the theory and is independent of the ultraviolet
regulator used for the quantum theory.  Fujikawa provided a new insight on
anomalies by showing that they arise due to the change of the fermion
measure under the corresponding transformation of the fermion
fields\cite{fujikawa} in the path integral method.  Chiral anomalies have a
deeper physical significance, as they relate the exact zero modes of the
Dirac operator to the nontrivial topological sectors of the gauge fields.
Consequently, the chiral anomaly in Quantum Chromodynamics(QCD) is thought
to give rise to $\eta'$ mass \cite{etamas}.   For the physically
interesting case of two massless flavor QCD ($N_f =2$), the order of the
chiral phase transition depends \cite{piswil} on the size of the coefficient of the chiral
anomaly term.  It is of second order, with critical exponents of
the $O(4)$ spin model, if the anomaly contribution is sizeable at finite
temperature.  One could expect a QCD-critical point in the $T-\mu$ plane
for light quarks in that case.  In view of this, it is important to
ascertain what change occurs in the anomaly in the presence of finite
temperature and densities.

\paragraph*{}   In this paper we address both the perturbative and
nonperturbative aspects of the chiral anomaly at finite temperature/density. In
Sec. I, we compute the triangle anomaly in the imaginary time formalism
of thermal  field theory. This method has the advantage that it can be
linked  to the weak coupling lattice calculations. Lattice QCD
deals with the imaginary time Euclidean propagators, and hence anomaly
calculation in the Euclidean space-time would be directly relevant for
numerical studies. In Sec. II, we extend  Fujikawa's analysis to finite
density in the continuum. We show that the anomaly equation arising due to
the change in the measure of the functional integrals under chiral
transformation of the fermion fields remains the same at nonzero densities
as well.
%in the basis of $D(0)$, the Dirac operator at zero density or in the basis
%of $D(\mu)$. 
We extend these considerations in Sec. III to the case of fermions with 
exact chiral invariance on the lattice.
%It is expected to continue remain true on the lattice as 
%the anomaly exists generically in gauge theories coupled to chiral fermions
%irrespective of the regulator we use. However the fermion operators with
%exact chiral symmetry on the lattice have explicit non-linear dependence on
%the chemical potential $\mu$. We show in section III that the correct
%anomaly relation cannot be obtained on the lattice for such operators at
%finite densities and discuss the possible consequences. 
We propose a lattice Dirac operator with  a term linear in the chemical 
potential $\mu$, i.e., similar to the continuum and also suggest a way
to get rid of the spurious divergences in the thermodynamic
quantities.  Its potential to  handle higher order terms in the 
Taylor expansion in chemical potential $\mu$ in full QCD is commented upon.

\section{Anomaly at $T=0$ and $\mu\neq0$ in continuum}

\subsection{Perturbative calculation}

\begin{figure}
 \begin{center}
 \includegraphics[scale=0.6]{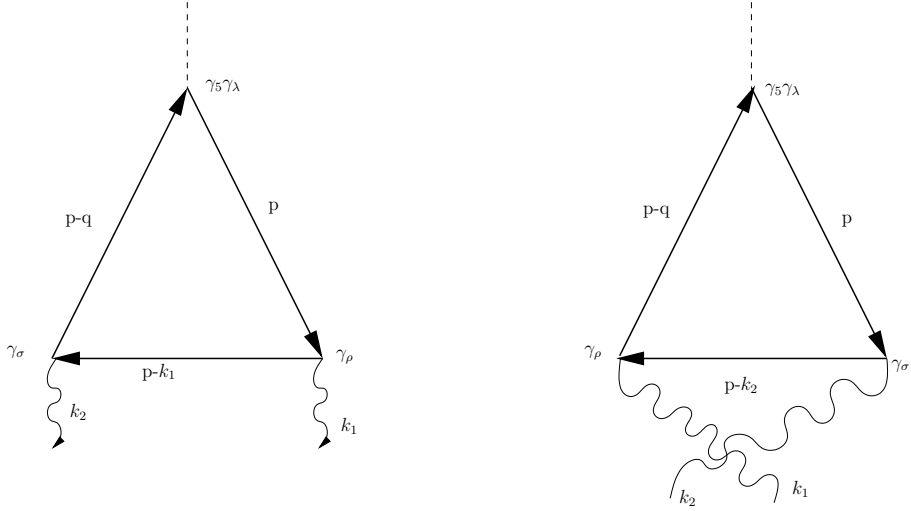}
\caption{ The ABJ triangle diagram(left panel) and its crossed counter part(right panel).}
\label{avv}
\end{center}
\end{figure}

In this section we calculate the expectation value of the gradient of flavor singlet axial
vector current of QCD perturbatively in the presence of finite fermion
density to check how the anomaly equation is affected in the presence of a
nonzero chemical potential. The lowest order diagrams are the ABJ triangle 
diagrams shown in Fig. \ref{avv}. It is well-known that the
higher order diagrams do not contribute to the anomaly equation  at zero
density, neither do other diagrams like the square and pentagon diagrams.
We therefore compute only the triangle diagrams at finite
density.  Our starting point is the QCD Lagrangian in the Euclidean space
with the finite number density term as defined in \cite{kapusta}.  In order
to maintain consistency with the lattice literature, we have however
chosen the Dirac gamma matrices to be Hermitian:

\begin{equation} \label{eqn:qcdl} 
\mathcal{L}=-\bar \psi({\not} D+m)\psi-\frac{1}{2}\text{Tr ~}F_{\alpha\beta}
F_{\alpha\beta} +\mu\bar \psi\gamma_4 \psi~,  
\end{equation} where
${\not}D=\gamma_{\nu}(\partial_{\nu}-igA^{a}_{\nu}T_{a})$ with $T_{a}$
being the generators of the SU(3) gauge group. The ghost terms are not
important in such a calculation as these do not directly couple to the
fermions.  The $\gamma_5=\gamma_1 \gamma_2 \gamma_3 \gamma_4$ is also
Hermitian in our case. The inverse free fermion propagator is seen to
acquire a $\mu$ dependence and become $[i{\not} p -m +\mu\gamma_4]$~.
In order to find out whether the chiral current 
$j_{\mu 5}=\bar\psi\gamma_\mu\gamma_5\psi$ for massless quarks
is conserved at finite density in one-loop perturbation theory, we compute
the quantum mechanical expectation value of the derivative of the chiral
current i.e. ,
\begin{equation}
\label{eqn:jmu5}
 \langle \partial_\mu j_{\mu,5}\rangle=-\frac{1}{2}\int d^4 x_1 d^4x_2\partial_\lambda\langle T \{j_{5,\lambda}(x)
 j_\rho(x_1)j_\sigma(x_2)\}\rangle A^{\rho}(x_1)A^{\sigma}(x_2)~.
\end{equation}
where the expectation value of the time ordered product of the three
currents at one-loop level is the axialvector-vector-vector (AVV) triangle diagram
shown in Fig. {\ref{avv}}. Any deviation of
this quantity from its classical value would give us the anomaly. Using the
Euclidean space Feynman rules, the amplitude of the AVV triangle diagram
can be computed. The crossed diagram is the one with the gluon legs
exchanged among the two vector (VV) vertices, and it corresponds to the process which is
quantum mechanically equally favored.

Denoting by $\Delta^{\lambda\rho\sigma}(k_1,k_2)$ the total amplitude and
contracting it with $q_{\lambda}$, Eq. (\ref{eqn:jmu5}) can be written in the 
momentum space for massless quarks as
\begin{eqnarray}
\nonumber
 q_{\lambda}\Delta^{\lambda\rho\sigma}&=& (- i)g^2 \text{tr}[T^a T^b]\int \frac{d^4 p}{(2\pi)^4}\text{Tr ~}\left[\gamma^5
\frac{1}{{\not} p-{\not}q-i\mu\gamma^4}
\gamma^\sigma\frac{1}{{\not} p-{\not}k_1 -i\mu\gamma^4}\gamma^\rho
-\gamma^5\frac{1}{{\not} p-i\mu\gamma^4}
\gamma^\sigma\frac{1}{{\not} p-{\not}k_1-i\mu\gamma^4}\gamma^\rho
\right.\\ 
&+&\left.\gamma^5
\frac{1}{{\not} p-{\not}q-i\mu\gamma^4}
\gamma^\rho\frac{1}{{\not} p-{\not}k_2-i\mu\gamma^4}\gamma^\sigma
-\gamma^5\frac{1}{{\not} p-i\mu\gamma^4}
\gamma^\rho\frac{1}{{\not} p-{\not}k_2-i\mu\gamma^4}\gamma^\sigma\right]~,
\end{eqnarray}
with the tr (Tr ) denoting trace over color (spin) indices.  Combining
further the first (second) term of the AVV diagram and the second (first)
term of the corresponding crossed diagram respectively,  we rewrite the
contracted amplitude in terms of functions $f_1(p,k_1)$ and $f_2(p,k_2)$,
\begin{equation}
\label{eqn:ano}
%\nonumber
 q_{\lambda}\Delta^{\lambda\rho\sigma}=(-i)~\text{tr}[T^a T^b]g^2\int \frac{d^4 p}{(2\pi)^4}\left[f_2(p-k_1,k_2)-f_2(p,k_2)+f_1(p-k_2,k_1)-f_1(p,k_1)
\right]~,
\end{equation}
where the function $f_1(p,k_1)$ is defined as,
\begin{eqnarray}
\label{eqn:fdef}
\nonumber
 f_1(p,k_1)&=&\text{Tr ~}\left[\gamma^5
\frac{{\not} p-i\mu\gamma_4}{(p_4-i\mu)^2+\vec p^2}
\gamma^\sigma\frac{{\not} p-{\not}k_1-i\mu\gamma_4}{(p_4-k_{14}-i\mu)^2+
(\vec p-\vec k_1)^2}\gamma^\rho\right]\\
&=-&\left[\frac{4\epsilon^{\alpha\sigma\beta\rho}p_\alpha k_{1\beta}-4i\mu\epsilon^{4\sigma\beta\rho}k_{1\beta}}{((p_4-i\mu)^2+\vec p^2)((p_4-k_{14}-i\mu)^2+
(\vec p-\vec k_1)^2)}\right]~,~\text{since Tr ~}[\gamma^5{\not} p\gamma^\sigma {\not} p\gamma^\rho]=0.
\end{eqnarray}
$f_2$ can be obtained by substituting $k_2$ for $k_1$ and interchanging
the indices $\rho$ and $\sigma$ in Eq. (5). We will use below a
common notation $f$ for denoting either in order to sketch
the proof further.
Although the numerator of Eq. (\ref{eqn:fdef}) has terms up to quadratic 
order in $\mu$, it should be noted that the $\mu^2$ terms 
are $ \sim\mu^2\text{Tr ~}[\gamma^5\gamma^4\gamma^\sigma\gamma^4\gamma^\rho]\sim 
\epsilon^{4\sigma 4 \rho}$ and therefore vanish. In order to further
evaluate the right-hand side of Eq. (\ref{eqn:ano}), we note that the integrals are 
linearly divergent and hence must be regulated by  introducing a cut-off scale,
$\Lambda$. This procedure must be carried out in a gauge invariant manner such
that the vector currents are conserved. In momentum space this amounts to
\begin{equation}
\label{eqn:gcon}
 k_{1\rho}\Delta^{\lambda\rho\sigma}(k_1,k_2)=k_{2\sigma}\Delta^{\lambda\rho\sigma}(k_1,k_2)=0~.
\end{equation}
We follow the usual text book \cite{zee} method to impose these conditions
above and compute the anomaly.  In order to highlight the differences due
to the $\mu \ne 0$ terms, we sketch below the evaluation of just the relevant
part of Eq.  (\ref{eqn:ano}).  Expanding the first term and combining
it with the second, we rewrite the first two integrals as,
\begin{equation}
\label{eqn:subtr}
\int \frac{d^4 p}{(2\pi)^4}\left[f(p-k_1,k_2)-f(p,k_2)\right]= \mathcal{L}t_{\Lambda\rightarrow \infty}\int_{0}^{\Lambda}\frac{d^4 p}{(2\pi)^4}
\left[ -k_{1\mu}\partial_\mu f+\frac{1}{2}k_{1\mu}k_{1\nu}\partial_\mu
\partial_\nu f +\mathcal{O}(k^3) \right]~.
\end{equation}
where the derivatives are in the momentum space. The first term of the above
 integrand can be written as a surface integral using Gauss law,
\begin{eqnarray}
\nonumber
 \mathcal{L}t_{\Lambda\rightarrow \infty}\int_{0}^{\Lambda}\frac{d^4 p}{(2\pi)^4}
k_{1\mu}\partial_\mu f(p,k_2)&=&\mathcal{L}t_{\Lambda\rightarrow \infty}
 \frac{k_{1\mu}\Lambda_\mu}{\Lambda}\frac{ f(\Lambda,k_2)2\pi^2\Lambda^3}{(2\pi)^4 }\\\nonumber
&\sim & \mathcal{L}t_{\Lambda\rightarrow \infty}
\left[\frac{4\epsilon^{\alpha\sigma\beta\rho}\frac{\Lambda_\alpha k_{1\mu} k_{2\beta}}{\Lambda}-\frac{4 i\mu}{\Lambda}\epsilon^{4\sigma\beta\rho}k_{1\mu}k_{2\beta}
}
{((1-i\frac{\mu}{\Lambda})^2+ 1)((1-\frac{k_{24}+i\mu}{\Lambda})^2+
(\hat\Lambda-\frac{\vec k_2}{\Lambda})^2)}\right]\frac{\Lambda_\mu\Lambda^3}
{8\pi^2\Lambda^4}\\
&=-&\frac{\epsilon^{\alpha\beta\sigma\rho} k_{1\alpha} k_{2\beta}}{8\pi^2}
\label{eqn:fmu}
\end{eqnarray}
where we uses the isotropy condition, $\Lambda_\nu \Lambda_\alpha /\Lambda^2
=g_{\nu\alpha}/4$. It is clear that the second term of the integrand in Eq.
(\ref{eqn:subtr}) when similarly integrated 
leads to the gradient of $f(p,k_2)$ at the Fermi surface of radius $\Lambda$, 
and therefore vanishes as $\mathcal{O}(\frac{1}{\Lambda})$.  Hence this term, and
the higher derivative terms, do not contribute in the limit when the cut-off is 
taken to infinity.  The other two terms of Eq. (\ref{eqn:ano}), as well as
the vector current conservation condition Eq. (\ref{eqn:gcon}), can be
similarly shown to be $\mu$ independent, leading to the canonical result
even for $\mu \ne 0$ :
\begin{equation}
 q_{\lambda}\Delta^{\lambda\rho\sigma}=-\text{tr}[T^a
T^b]\frac{ig^2}{2\pi^2}\epsilon^{\alpha\beta\sigma\rho}k_{1\alpha}
k_{2\beta}~.
\end{equation}

We have thus shown explicitly that the anomaly equation has no 
corrections due to nonzero $\mu$ or, equivalently, at nonzero finite density.
It is easy to generalize the same computation to nonzero temperatures. At 
finite temperature, the temporal part of the momentum gets quantized as the
well-known Matsubara frequencies : $p_4=\frac{(2n+1)\pi}{\beta}$. 
Correspondingly, $\int_{-\infty}^{\infty} \frac{dp_4}{2\pi}$ gets replaced
by $ \frac{1}{\beta}\sum_n$,  where $n=\pm 1,\pm 2,...,\pm \infty$. The sum
over discrete energy eigenvalues, can as usual, be split as a zero
temperature contribution along with the finite temperature contributions
weighted by the Fermi-Dirac distribution functions for the particles and
antiparticles.  Note that the finite temperature contributions will fall off to 
zero in the ultraviolet limit because these are regulated by the distribution 
functions. Thus,
\begin{eqnarray}
 \label{eqn:fano}
\nonumber
&&\int \frac{d^3 \vec p}{(2\pi)^3}\left[k_1^i\partial_i\left[
 f(|\vec p|) \left(\frac{1}{e^{\beta \left(|\vec p|-\mu\right)}+1}+\frac{1}
{e^{\beta\left(|\vec p|+\mu\right)}+1}\right)\right]+\{\rho,k_1\leftrightarrow\sigma,k_2\}
\right]\\
&&=\mathcal{L}t_{|\vec p|\rightarrow \infty}\frac{4\pi |\vec p|}{(2\pi)^3}
\left[(\vec k_1\cdot \vec p)
 f(|\vec p|) \left(\frac{1}{e^{\beta\left(|\vec p|-\mu\right)}+1}+\frac{1}
{e^{\beta \left(|\vec p|+\mu\right)}+1}\right)+\{\rho,k_1\leftrightarrow\sigma,k_2\}
\right]\longrightarrow0
\end{eqnarray}
Such perturbative calculations of the ABJ anomaly were reported earlier in
the real time formalism at finite temperature and at both zero \cite{ito}
and nonzero \cite{su,sn} fermion densities as well as for finite density in
Minkowski space-time \cite{sannino}. We have shown above that these
calculations are possible using the imaginary time formalism as well.
An imaginary time calculation is useful as this can be generalized to
weak coupling calculations in lattice gauge theory.

\subsection{Nonperturbative calculation}
The chiral anomaly in the path integral formalism can also be looked upon as
arising due to the change of the measure under chiral transformation of the
fermion fields\cite{fujikawa}.  In this section, Fujikawa's method of anomaly
calculation in the path integral formalism, at zero temperature and zero
fermion density, is extended to the finite fermion density case.  But
before analyzing the finite density problem, the method for $\mu=0$ is
summarized to point out the differences that would arise in the finite density
case. The partition function for massless fermions interacting with $SU(N)$
gauge theory can be written in Euclidean space as
\begin{equation}
\text{Z}=\int \mathcal{D}\bar \psi \mathcal{D}\psi [\mathcal{D}A_{\nu}]
\rm{e}^{-\int d^{4} x ~\bar \psi{\not}D\psi-S_{YM}}
=\int \mathcal{D}\bar \psi \mathcal{D}\psi [\mathcal{D}A_{\nu}]\rm{e}^{-S}
\end{equation}
where $S_{YM}=1/2\int d^4 x\left[\text{Tr ~}F_{\alpha\beta}(x)F_{\alpha\beta}(x)
+1/\xi(f^aA_\mu^a)^2\right]$
 is the free Yang-Mills action with appropriate gauge fixing $f^a A_\mu^a=0$.
 The action for the ghost term is included within the gauge field
 measure and hence denoted within square brackets. This is justified
 since we are interested in the change of the fermion fields under chiral
transformations and the ghost fields do not interact with the fermions. 
 Under the infinitesimal local chiral
 transformation of the fermion fields, given by
\begin{equation}
\label{eqn:ct}
\delta \psi(x) = i\alpha (x) \gamma_5 \psi(x)  ~~~{\rm and} ~~~  
\delta \bar \psi(x) = i\alpha(x) \bar \psi(x) \gamma_5 ~,~
\end{equation}
the action changes as $S\rightarrow S-i\int d^4
x~\alpha(x)\partial_{\nu}j^{\nu}_5$.  
The measure changes as a result of the transformation of the fermion
fields. The change of measure is,
\begin{equation}
\label{eqn:cj}
 \mathcal{D}\bar \psi^{'} \mathcal{D}\psi^{'}=\mathcal{D}\bar \psi \mathcal{D}\psi
\text{Det}\vert\frac{\partial(\bar \psi^{'},\psi^{'})}{\partial(\bar \psi,\psi)}\vert
=\mathcal{D}\bar \psi \mathcal{D}\psi\rm{e}^{-2i\int d^4x~ \alpha(x) \text{Tr}\gamma_5 }
\end{equation}
where Tr stands for the trace over the spin and the color space. This trace can be
computed using the eigenvectors of the operator ${\not}D$, since it  is
an anti-Hermitian operator.  It has purely imaginary eigenvalues and  the
corresponding eigenvectors form  a complete orthonormal basis.  Splitting
the trace computation into two parts, the trace over the nonzero eigenvalues
can be done easily as follows.  Since $\{\gamma_5,{\not}D\}=0$, for every
eigenvector $\phi_{m}$ with nonzero eigenvalue $\lambda_m\neq0$, there is a
corresponding eigenvector $\gamma_5\phi_{m}$ with eigenvalue $-\lambda_m$.
Thus for each finite $\lambda_m$ ,
$\phi^{\pm}_{m}=\phi_{m}\pm\gamma_5\phi_{m}$ are eigenvectors of
$\gamma_5$ with eigenvalues $\pm1$. Since trace is independent of the basis
vectors we can also compute the trace of $\gamma_5$ in the $\phi^{\pm}_{m}$
basis. One obtains zero as the result since there are equal number of
$\phi^{\pm}_m$ respectively.  For the zero eigenmodes, ${\not}D$ and $\gamma_5$
commute hence each zero mode has a definite chirality, leading to a +1
contribution for those with $\gamma_5 \phi_n = \phi_n$ and a -1 for the opposite
chirality.  Hence the complete evaluation of the trace gets a nonzero
contribution corresponding to the difference between number of the two
chiralities: 
\begin{equation}
\label{eqn:ana}
\text{Tr} \gamma_5=\sum_n\phi_{n}^{\dagger}\gamma_5\phi_{n}=n_{+}-n_{-}.
\end{equation}
\subsection*{Chiral Jacobian in the presence of $\mu$}
The presence of finite chemical potential, $\mu$, in the action can be
described as an effective change of the Dirac operator from ${\not}D$ to
${\not}D-\mu\gamma_4={\not}D(\mu)$.  Under the chiral transformation given
in Eq. (\ref{eqn:ct}) the action still remains invariant as in the zero
density case.  This is due to the fact that the $\mu$ dependent term of the
action anticommutes with $\gamma_5$: $\{\gamma_5,\mu\gamma_4\}=0$.  Under
the transformations given in Eq. (\ref{eqn:ct}) the fermion measure changes
again by the same Jacobian factor $\text{Tr}\gamma_5$.  The
corresponding $\text{Tr}\gamma_5$ is now evaluated in the space of
eigenvectors of the new Dirac operator ${\not}D(\mu)$.  This is
because the measure is defined by the complete set of states of the Dirac
operator which appears in the action. Although ${\not}D(\mu)$ has both an
anti-Hermitian and a Hermitian term, it is still 
diagonalizable.   Consider an eigenvector $\phi_m$ of ${\not}D(0)$ with an
eigenvalue $\lambda_m$.  Let us define two new vectors, $\zeta_m$ and
$\upsilon_m$ as follows:

\begin{equation}
 \label{eqn:chdiag}
\zeta_m(\mathbf x, \tau)=\rm{e}^{\mu\tau} \phi_m ( \mathbf x ,
\tau)~~,~~\upsilon^{\dagger}_m ( \mathbf x, \tau)=\phi^{\dagger}_m ( \mathbf x , 
\tau)\rm{e}^{-\mu\tau}~.
\end{equation}
It is easy to check that $\zeta_m$ is the eigenvector of ${\not}D(\mu)$
with the same (purely imaginary) eigenvalue $\lambda_m$,
\begin{equation}
 {\not}D(\mu)\zeta_m=\lambda_m\zeta_m~,~
\end{equation}
and  $\upsilon^{\dagger}_m$ is the eigenvector of ${\not}D(\mu)^\dagger$
with the eigenvalue $\lambda_m^* = -\lambda_m$,
\begin{equation}
\upsilon^{\dagger}_m{\not}D^{\dagger}(\mu)=-\lambda_m\upsilon^{\dagger}_m.
\end{equation}

Note that the sets of eigenvectors $\{\zeta\}$ and $\{\upsilon\}$ are in
one-to-one correspondence with the complete set $\{\phi\}$. Using the
completeness relation for the latter, 
\begin{equation}
\sum_m \int\phi_m(\mathbf x, \tau) \phi_m^{\dagger}(\mathbf x, \tau)~d^4 x=\mathbf{I}~,~
\end{equation}
where $\mathbf{I}$ denotes the identity matrix,  we note that
\begin{equation}
\label{eqn:chcmp}
\sum_m \int\zeta_m(\mathbf x, \tau) \upsilon_m^{\dagger}(\mathbf x, \tau)~d^4 x=
\sum_m \int\phi_m(\mathbf x, \tau)\rm{e}^{\mu\tau}\rm{e}^{-\mu\tau} 
\phi_m^{\dagger}(\mathbf x, \tau)~d^4 x =\mathbf{I}~.
\end{equation}
Moreover,  $\{\zeta\}$ and $\{\upsilon\}$ satisfy the following normality 
condition,
\begin{equation}
\label{eqn:chnrm}
\int \upsilon_m^{\dagger }(\mathbf x, \tau)\zeta_m(\mathbf x, \tau)~d^4 x=~\int \phi_m^{\dagger}\rm{e}^{-\mu\tau}\rm{e}^{\mu\tau}\phi_m ~d^4 x
=~\int\phi_m^{\dagger}(\mathbf x, \tau)\phi_m(\mathbf x, \tau)~d^4 x = 1 ~,~
\end{equation}
leading to 
\begin{equation}
 \upsilon_m^{\dagger}(\mathbf{x}, \tau) \gamma_5 \zeta_m(\mathbf{x},
\tau)=\phi_m^{\dagger}\rm{e}^{-\mu\tau} \gamma_5 \rm{e}^{\mu\tau}\phi_m
=~\phi_m^{\dagger}(\mathbf{x}, \tau)\gamma_5 \phi_m(\mathbf{x}, \tau) ~,~
\end{equation}
Using these eigenvector spaces of ${\not}D(\mu)$, the calculation of 
$\text{Tr}\gamma_5$ goes through in the same way as for ${\not}D(0)$ above. 
Since the new operator still anticommutes with $\gamma_5$ i.e
$\{\gamma_5,{\not}D(\mu)\}=0$,  for each eigenvector $\zeta_m$
with eigenvalue $\lambda_m$ there is an eigenvector
$\gamma_5\zeta_m$ with the eigenvalue $-\lambda_m$.  Thus the trace of
$\gamma_5$ is zero for all nonzero $\lambda_m$.  
In the basis of the zero modes of ${\not}D(\mu)$, given by $\zeta_n$ 
and $\upsilon^\dagger_n$, the change in the fermion measure is given as,
\begin{equation}
\label{eqn:chJ}
\text{Tr} \gamma_5=\sum_n\upsilon_{n}^{\dagger}\gamma_5\zeta_{n}
=\sum_n\phi_{n}^{\dagger }\rm{e}^{-\mu\tau}\gamma_5\rm{e}^{\mu\tau}\phi_{n}=n_{+}-n_{-}.
\end{equation}
Thus the change in the fermion measure due to the chiral transformations is the
same as in the zero density case with no additional $\mu$ dependent terms.
Hence the anomaly is unaffected in the presence of $\mu$.  Some remarks on the
proof may be in order.  The definition of the vectors  $\zeta_m$ and
$\upsilon_m$ in Eq. (\ref{eqn:chdiag}) assumes that neither $\mu$ nor $\tau$ is
infinite. The same assumption is also utilized in various steps in 
Eqs. (\ref{eqn:chcmp})-(\ref{eqn:chJ}).  Clearly at strictly zero temperature,
this is not tenable.  However, an infinitesimally small temperature suffices
for the proof to go through.  Moreover, since the result is finally
$\mu$-independent, we expect the result to be valid at zero temperature,
although our proof is valid only in the limit of zero temperature.  
%, as expected na{\i}vely, since it is an ultraviolet effect.   
The scaling
of the eigenvectors, including the chiral zero modes, by the exp$( \pm \mu
\tau)$ factors  can be related to a nonunitary
transformation of the fermion fields in the QCD action in the presence of
$\mu$, given by
\begin{equation}
 \label{eqn:chinu}
\psi^{'}(\mathbf x, \tau)=\rm{e}^{\mu\tau} \psi ( \mathbf x , \tau)~~,~~
\bar \psi^{'}( \mathbf x, \tau)= \bar \psi( \mathbf x, \tau)\rm{e}^{-\mu\tau}~,
\end{equation}
which makes the action $\mu$-independent:
\begin{equation}
 S =\int d^4 x~ \bar\psi^{'}[{\not}D-\mu\gamma_4]\psi^{'}
=\int d^4 x~ \bar\psi\rm{e}^{-\mu\tau}~[{\not}D-\mu\gamma_4]~\rm{e}^{\mu\tau}\psi
=\int d^4 x~\bar\psi~{\not}D~\psi~.
\end{equation}

Note that the fields $\psi$ and $\bar\psi$ at the same space-time point
scale differently in the transformation in Eq. (\ref{eqn:chinu}) which is
permissible \cite{pm} in the Euclidean field theory since they are mutually
independent fields.  Let us also emphasize that the transformation in 
Eq. (\ref{eqn:chinu}) is not unitary and thus not physical.  Indeed, it
merely relates the actions in two different physical situations of zero
and nonzero $\mu$.  One clearly cannot employ it in the evaluation of  
the partition function due to its nonunitary nature.  We have shown above 
that the transformation suggests how to extend the cancellation argument 
for nonzero eigenvalues of the Dirac operator for $\mu=0$ to the nonzero $\mu$ 
case as well and is thus useful.  Furthermore, since the transformation 
commutes with both flavor singlet and nonsinglet chiral transformations, 
employing it as a prescription to introduce the chemical potential will 
necessarily lead to a $\mu$ dependent action which has the same chiral 
invariance as for $\mu=0$.  Whether this  way to introduce the chemical 
potential in any theory is the only way to do so without affecting its 
chiral invariance would be interesting to explore; we conjecture that this 
is the case. 

\section{Anomaly on the lattice at finite density}

The above discussion of the anomaly in the continuum suggests a way to
introduce the chemical potential on the lattice.  By preserving the
transformation (\ref{eqn:chinu}) on the lattice, one may expect to maintain
the anomaly to remain  $\mu$ independent on the lattice as well.  Let us
consider the na{\i}ve massless fermion action on the lattice,
\begin{equation}
S=-\sum_{x,y}\bar\psi_x\left[U_4^{\dagger}(x-\hat{4})\frac{\gamma_4}{2 }
\delta_{x,y+\hat{4}}-U_4(x)\frac{\gamma_4}{2 }\delta_{x,y-\hat{4}}+
\sum_{i=1}^{3}\left(U_i^{\dagger}(x-\hat{i})\frac{\gamma_i}{2 }
\delta_{x,y+\hat{i}}-U_i(x)\frac{\gamma_i}{2 }\delta_{x,y-\hat{i}}\right)\right]\psi_y~. 
\end{equation}
Replacing the $\psi$ and $\bar \psi$ fields in the above action by $\psi'$
and $\bar \psi'$ respectively, using the lattice analogue of the
transformation (\ref{eqn:chinu}), we indeed obtain
a fermionic action on the lattice at finite density,
\begin{equation}
\label{eqn:naivmu}
 S=-\sum_{x,y}\bar\psi^{'}_x\left[\rm{e}^{-\mu a_4}U_4^{\dagger}(x-\hat{4})\frac{\gamma_4}{2 }
\delta_{x,y+\hat{4}}-\rm{e}^{\mu a_4}U_4(x)\frac{\gamma_4}{2 }\delta_{x,y-\hat{4}}+
\sum_{i=1}^{3}\left(U_i^{\dagger}(x-\hat{i})\frac{\gamma_i}{2 }
\delta_{x,y+\hat{i}}-U_i(x)\frac{\gamma_i}{2 }\delta_{x,y-\hat{i}}\right)\right]\psi^{'}_y~.
\end{equation}
with $a_4$ being the lattice spacing in the temporal direction.
Unfortunately, the infamous fermion doubling problem is related to the fact
that the anomaly on the lattice is canceled exactly for such na{\i}ve
fermions.  The ``no-go'' theorem of Nielsen and Ninomiya  \cite{nini}
states that it is impossible to construct lattice Dirac operators which
simultaneously satisfy Hermiticity, and locality and have chiral symmetry while
being free of the ``doublers''.  The commonly used fermions on the lattice,
like the Wilson and the Kogut-Susskind fermions do not have 
$U_A(1)$ chiral symmetry, and so there is no anomaly to speak of.
Nevertheless, we note that a similar transformation for such fermions does lead
to the action popularly used for nonzero chemical potential\cite{hk,kogut}.

Recently, Neuberger \cite{NeuNar} constructed a fermion 
operator $D_{ov}$, commonly known as the overlap operator, which has exact
chiral symmetry and satisfies the Ginsparg and Wilson\cite{gw} relation, 
%the modified chirality relation on the lattice given as, Ginsparg and
%Wilson\cite{gw} suggested a way to evade the ``no-go'' theorem: they
%prescribed a modified chiral invariance condition for the Dirac operator
%on the lattice such that the doublers are absent and it is possible to
%simulate a chiral
%fermion on the lattice. Narayanan and Neuberger constructed a fermion
%operator with exact chiral symmetry on the lattice in the massless limit,
%known as the overlap fermions\cite{NeuNar}.  These fermions with exact
%chiral symmetry on the lattice satisfy the Ginsparg-Wilson relation given
%as
\begin{equation}
\label{eqn:neudir}
 \{ \gamma_5, D_{ov} \} = D_{ov}\gamma_5 D_{ov}~~ \text{with} ~~ D_{ov}=1
+\gamma_5\epsilon(\gamma_5 D_{W})~.~
\end{equation}
Here $\epsilon$ is the sign function and $D_{W}$ is the canonical 
Wilson-Dirac operator with a parameter $M$,
\begin{eqnarray}
 \label{eqn:wilopnomu}
D_W(x,y)=\left(4-M\right)\delta_{x,y}-
\sum_{i=1}^{4}\left(U_i^{\dagger}(x-\hat{i})\frac{1+\gamma_i}{2 }
\delta_{x,y+\hat{i}}+U_i(x)\frac{1-\gamma_i}{2 }\delta_{x,y-\hat{i}}\right)~.
\end{eqnarray}
The value of the parameter $M$ is constrained to lie between 0 and 2 for
simulating a one flavor quark on the lattice. 
The overlap fermion action is invariant under the following chiral 
transformation, as shown by Luscher \cite{luscher}, 
\begin{equation}
\label{eqn:chrl}
\delta \psi = \alpha  \gamma_5(1 - \frac{1}{2}D_{ov}) \psi  ~~~{\rm and} ~~~  
\delta \bar \psi = \alpha \bar \psi (1 - \frac{1}{2}D_{ov})\gamma_5 ~.~
\end{equation}
At zero temperature and density, the change in the measure computed on the
lattice due to the Luscher transformations was shown to be related to the
index of the fermion operator \cite{hln,luscher,fujikawal} ,
 \begin{equation}
 \text{Tr} ~[2\gamma_5(1-\frac{1}{2} D_{ov})]=-\text{Tr} ~(\gamma_5 D_{ov})=n_{+}-n_{-}
=~2 ~\text{Index}D_{ov}~,
\end{equation}
where $n_{\pm}$ are right and left handed fermion zero modes respectively.

Bloch and Wettig \cite{wettig} proposed a method to incorporate the chemical 
potential in the overlap operator.  It consisted of i) multiplying $U_4$
[$U^\dag_4$] by exp($\mu a_4$)[exp(-$\mu a_4$)] in the $D_W$ in 
Eq. (\ref{eqn:wilopnomu}) and ii) extending the definition of the sign function
for the resultant complex matrix.   The $D_{ov}(\mu)$ also satisfied the
 Ginsparg-Wilson relation :
\begin{equation}
  \{ \gamma_5, D_{ov}(\mu) \} = D_{ov}(\mu)\gamma_5 D_{ov}(\mu).~~\text{with}~~D_{ov}(\mu)=1+\gamma_5\epsilon(\gamma_5 D_{W}(\mu))~.
\end{equation}
It should be noted that the resultant action does not have the property of
eliminating the $\mu$-dependence by any transformation like Eq.
(\ref{eqn:chinu}) due to the nonlocal nature of $D_{ov}$. 
 
As we pointed out\cite{bgs} earlier though, the action  $S=\sum_ {x,y} 
~\bar\psi_x[ D_{ov}(\mu)]_{xy}\psi_y$ is not invariant under Luscher's chiral 
transformations of Eq. (\ref{eqn:chrl}).  Indeed, its variation is easily
seen to be
\begin{equation} 
\nonumber 
\delta S =\frac{a \alpha}{2} \sum_{x,y} \bar \psi_x 
\big[2 D_{ov}(\mu) \gamma_5 D_{ov}(\mu) - 
D_{ov}(0) \gamma_5 D_{ov} (\mu) - D_{ov}(\mu)
\gamma_5 D_{ov}(0) \big]_{xy} \psi_y \ne 0 ~.~ 
\end{equation} 

The chiral symmetry violation is of the $\mathcal{O}(a)$ and
hence the symmetry is restored in the continuum limit.  One may 
alternatively propose modified chiral transformations, 
\begin{equation}
\label{eqn:chrlbw}
\delta \psi = \alpha  \gamma_5(1 -\frac{1}{2}D_{ov}(\mu)) \psi  ~~~{\rm and} ~~~  
\delta \bar \psi = \alpha \bar \psi (1 - \frac{1}{2}D_{ov}(\mu))\gamma_5 ~,~
\end{equation}
which will ensure $\delta S = 0$.  In that case, the anomaly equation
$-\text{Tr} ~(\gamma_5 D_{ov}(\mu))=2 ~\text{Index}D_{ov}(\mu)$ is valid
\cite{wettig} on the lattice even in the presence of $\mu$, since the
fermion measure changes under these transformations by a Jacobian factor
$\text{Tr} ~[2\gamma_5(1-1/2 D_{ov}(\mu))]$.  Note, however, that the relevant zero
modes are now those of the $D_{ov}(\mu)$, and thus $\mu$ dependent, in
contrast to our continuum result of the previous section.

Furthermore, altering the symmetry transformations as above has undesirable
physical consequences, as discussed in detail in \cite{bgsL}.  Let us briefly
outline here the main points. Non-Hermiticity of $\gamma_5 D_{ov}(\mu)$
makes the transformations nonunitary.  The symmetry transformations should
not depend on the intensive thermodynamic quantity $\mu$, which is a tunable
parameter of the physical system.  The symmetry group itself changes with
$\mu$, leaving no physical order parameter which will
characterize the chiral phase transition as a function of $\mu$.  In
contrast, the chiral symmetry group remains the same at nonzero temperature
(and zero density), allowing us to infer that vanishing of the chiral
condensate would correspond to restoration of the symmetry for the vacuum.

\subsection{A simple proposal}

It is well-known that the overlap fermion operator can be obtained
\cite{Neu2,eh} from the five dimensional domain wall fermions in the limit of
infinite extent of the fifth dimension.  The Bloch-Wettig proposal above
was also shown to arise \cite{bloch} in this way.  It turns out that the
chemical potential, $\mu$  enters in their action then as the Lagrange
multiplier for the number of fermions on {\em each} slice of the fifth
dimension.  This means that all the unphysical ``bulk'' modes are
considered with the same weightage in the partition function as the zero
modes which eventually correspond to the massless quarks in four
dimensions. The subsequent cancellation of the bulk contributions using
Pauli-Villars fields to single out the contribution  of a single chiral
fermion thus becomes $\mu$ dependent on the lattice.  Motivated by this
physical fact, we propose to introduce the chemical potential only to count
the fermion confined to the domain wall. Integrating out the fermions in
the fifth dimension, one is led to the following action, which one
would have written down naively to add a number density term :

\begin{equation}
\label{eqn:wilopmu}
 D_{ov}(\hat\mu)_{xy}=(D_{ov})_{xy} 
-\frac{a \hat\mu}{2a_4 ~M}\left[
(\gamma_4+1)U^{\dagger}_4(y)\delta_{x,y+\hat{4}}-(1-\gamma_4)U_4(x)\delta_{x,y-\hat{4}}\right]~.
\end{equation}

Here $D_{ov}$ is the same Neuberger-Dirac operator of
Eq. (\ref{eqn:neudir}), and $\hat\mu=\mu a_4$ is the chemical potential in
lattice units.  As usual, the volume of the system is $V=N^3a^3$ and
the temperature is $T=1/(N_Ta_4)$  on a $N^3\times N_T$ lattice with
lattice spacings $a$ and $a_4$ in spatial and temporal directions
respectively.  The term containing the chemical potential is not unique.
Improved density operators could be used for faster approach to the
continuum limit, e.g., addition of three-link terms.  We could have chosen
$\hat\mu/s$ instead of $\hat\mu/M$ as the multiplying factor for the
conserved number part. All such choices of actions are constrained by the
fact that these have the correct continuum limit. However the finite
lattice spacing errors in each of these operators would be different and we
comment below on how they may affect the numerical simulations.

Note that our proposal, too, will break  exact chiral invariance at the same
$\mathcal{O}(a)$ as the Bloch-Wettig proposal.  As a result, the anomaly
equation on the lattice will get $\mu$ -dependent corrections anyway. A significant
difference may be the fact that the change in the measure is $\mu$ independent
for our proposal, as in the case of the
continuum.  We persist with it in the following, nevertheless,
as it is simpler and easier to implement. Principally, this is due to the
fact that one does not have to compute the sign function of a non-Hermitian 
matrix, with its inherent ambiguities, as in the Bloch-Wettig way of
incorporating the chemical potential. The non-Hermitian sign function is
numerically also more expensive to simulate for the full interacting case,
whenever that becomes more practical.

For noninteracting fermions the $U_{\mu}=1$ and the above Neuberger-Dirac
operator with the chemical potential term can be diagonalized in 
momentum space in terms of the functions,
\begin{eqnarray}
\label{eqn:hdef}
\nonumber
h_j&=&-\sin a p_j~,~h_4=-\frac{a}{a_4}  \sin(a_4p_4)~,~\\
h_5&=&M-\sum_{j=1}^{3}(1-\cos a p_j)
-\frac{a}{a_4}(1- \cos (a_4 p_4))~,s=\sqrt{\sum_{j=1}^{3}h_j^2+h_4^2+h_5^2}
\end{eqnarray}
such that $D_{ov}(\hat\mu)$ can be written as,
\begin{equation}
\label{eqn:ovmod}
D_{ov}(\vec p,p_4,\hat\mu) =1-\sum_{i=1}^{4} i\gamma_i \frac{h_i}{s}-\frac{h_5}{s}-
\frac{a \hat\mu}{a_4 M}\left[\gamma_4 \cos(a_4 p_4)-i \sin(a_4 p_4)\right]~.
\end{equation}
%\begin{equation}
%\label{eqn:ovmod}
%D_{ov}(\vec p,p_4) =1-\sum_{i=1}^{4} i\gamma_i \frac{h_i}{s}-\frac{h_5}{s}-
%\frac{a}{a_4 M}\left[i\gamma_4 \sin(a_4 p_4-i\hat\mu)-i\gamma_4 \sin(a_4 p_4)-\cos(a_4 p_4-i\hat\mu)+\cos(a_4 p_4)\right]~.
%\end{equation}
%In the previous equation we have rewritten $K$ and $L$ in terms of $R$ and $\theta$ such that,
%\begin{equation}
 %\frac{K(\hat{\mu})+L(\hat{\mu})}{2}=\cosh \hat\mu ~,~~
%\frac{K(\hat{\mu})-L(\hat{\mu})}{2}=\sinh \hat\mu~,
%\end{equation}
To study thermodynamics of fermions one has to necessarily take antiperiodic
boundary conditions along the temporal direction. Assuming periodic boundary
conditions along the spatial directions we obtain
\begin{eqnarray}
\nonumber
ap_j &=& \frac{2n_j\pi}{N}~,~n_j=0,..,(N-1),~j=1,~2,~3 ~{\rm and}\\
ap_4 &=&\omega_n= \frac{(2n+1)\pi}{N_T}~,~n=0,..,(N_T-1)~,
\end{eqnarray}
where $\omega_n$ are the Matsubara frequencies.  The operator given by Eq.
(\ref{eqn:ovmod}) can be shown to have correct continuum limit. The number
density can be calculated at zero temperature by the contour integral
method as was discussed for the Bloch-Wettig version of the overlap
fermions at finite $\mu$ in \cite{bgs}.  The major difference one finds is
the expected $\mu/a^2$-divergence ($\mu^2/a^2$-divergence) in the number
(energy) density in the continuum limit of $a \to 0$.  What is perhaps not
widely appreciated from such calculations is that the leading term,
corresponding to the Stefan-Boltzmann limit, also changes by a {\em finite}
computable part.   In the next section, we show through numerical
evaluations of the sums, how one can deal with these problems.

\subsection{Numerical Results}

We compute two thermodynamic quantities of relevance to the above
discussion as well as to the heavy-ion collision experiments: the change
in the energy density due to the chemical potential,
$\Delta\varepsilon(\mu,T)=\varepsilon(\mu,T)-\varepsilon(0,T)$ and the
quark number susceptibility at zero chemical potential, $\chi(0)$.
These thermodynamic quantities are computed by
taking appropriate derivatives of the partition function $Z= \det
D_{ov}$,
\begin{equation} 
\chi(0)=\frac{1}{N^3 a^2 N_T}\left(\frac{\partial^2 \ln \det D_{ov}}
{\partial \hat{\mu}^2 }\right)_{a_4,\hat\mu\rightarrow0,a_4=a}~,~\varepsilon
(\hat\mu)=-\frac{1}{N^3 a^3 N_T}\left( \frac{\partial \ln \det D_{ov}}
{\partial a_4 }\right)_{\hat\mu N_T,~a_4=a} 
\end{equation} 
The quantities computed on the lattice are expected to have a
$\Lambda^2\sim 1/a^2$ dependence on the lattice.  In order to eliminate
these spurious $\Lambda^2$ terms, we follow the same prescription which was
used for the energy density computation at zero temperature (which diverges
as $\Lambda^4$ ).  We compute these thermodynamic quantities at zero
temperature and subtract them from the corresponding values computed on the
lattice at nonzero temperatures.  The zero temperature values were computed
numerically on a lattice with a very large temporal extent $N_T$ and
fixed $a_4$ such that $T=1/(N_T a_4)\rightarrow0$.  The Matsubara
frequencies then become continuous and hence could be integrated upon
numerically.
\begin{figure}
\begin{center}
 \includegraphics[scale=0.6]{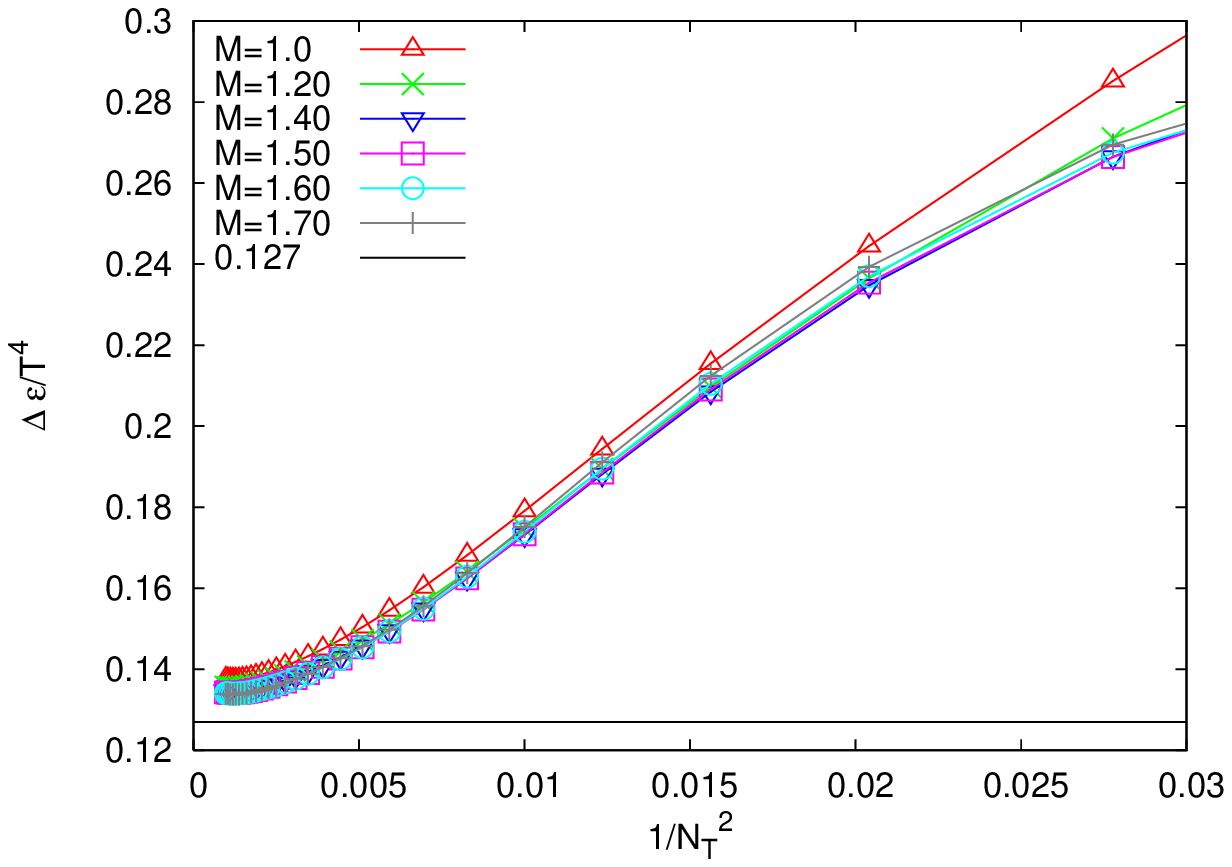}
\includegraphics[scale=0.6]{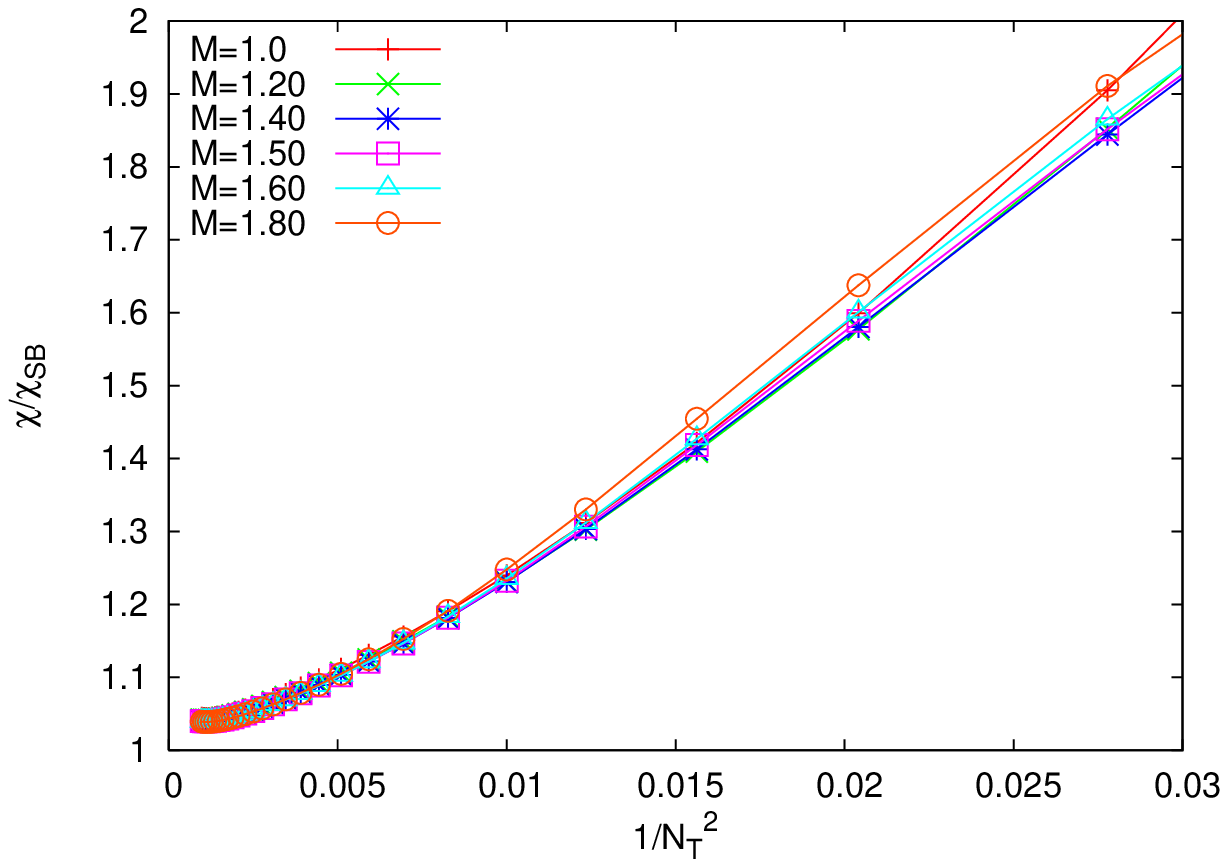}
\caption{ The energy density(left panel) and quark number susceptibility (right panel)as a function of $1/N_T^2$ for $M$ values as indicated for $\zeta =4$.}
\label{ovmu}
\end{center}
\end{figure}

Fig. \ref{ovmu} displays the subtracted results for
$\Delta\varepsilon(\mu,T)$ for $r = \mu/T = \hat\mu N_T = 0.5$ and
$\chi(0)$.  The former is displayed in units of $T^4$ and has the value
0.127 for $r = 0.5$ in the continuum limit, while the latter is
normalized to the ideal gas value ($T^2/3$).  The $M$ values are as
indicated along the symbol used.  The subtraction constants had to
be computed separately for energy density and susceptibility.
From a comparison of the plots with the
corresponding ones \cite{bgs} for the Bloch-Wettig case, we find that
\begin{itemize}
\item there are no leftover effects of divergences after the zero 
temperature subtraction,
\item there are no oscillations for odd-even values of $N_T$,
\item the M-dependence is much less pronounced, and
\item the scaling towards the continuum value is linear with the
possibility of an easier extrapolation.
\end{itemize}

We also computed the susceptibility using the Wilson fermions and
compared the results with those above.  We found that for $N_T=6$ the 
cut-off effects of the Wilson operator are about $21\%$ larger than the 
$M=1.60$ overlap result shown in the right panel of Fig. (\ref{ovmu}). 
The difference reduces to about $3\%$ at $N_T=10$. Beyond $N_T=10$, the 
approach to the continuum limit is almost identical for both the operators.
The Wilson fermions have no chiral symmetry even for $\mu=0$, which
may make them less favored for the QCD critical point searches which
are pivoted around the $\mu=0$ transition.

\begin{figure}
\begin{center}
\includegraphics[scale=0.65]{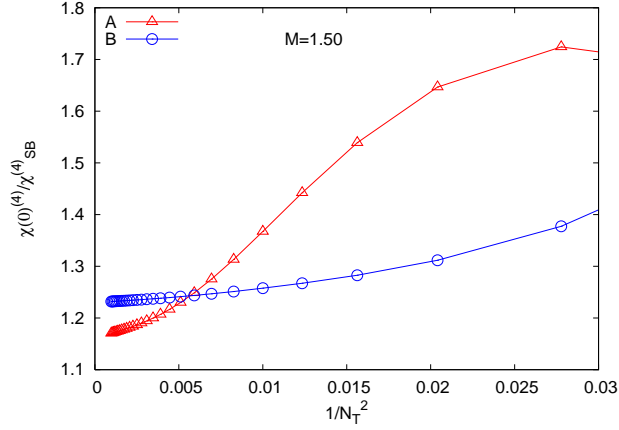}
\caption{ The variation of the ratio of the fourth order susceptibility and the corresponding continuum value as a function of $1/N_T^2$ for $\zeta =4,~M=1.5$ for the A) $\hat\mu/s$ and 
B)$\hat\mu/M$ ways of incorporating the chemical potential.}
\label{sus4}
\end{center}
\end{figure}

We have also checked that there are no other divergent terms of the form
$\mathcal{O}(a^{-n})$ with $n>2$ in the number density, by calculating
the fourth-order susceptibility since odd orders of susceptibilities vanish
at $\mu=0$. At zero chemical potential, the fourth-order susceptibility
is given by,
\begin{equation}
 \chi^{(4)}(0)=\frac{1}{N^3 N_T}\left(\frac{\partial^4 \ln \det D_{ov}}{\partial
\hat{\mu}^4 }\right)_{a_4,\hat\mu\rightarrow0}
\end{equation}
A term $\mathcal{O}(a^{-4})$ in the number density will show up as a
divergence in this susceptibility, and will need a subtraction too.
From Fig. (\ref{sus4}), where we display our results for
$\chi^{(4)}(0)$ for $M=1.5$, we can conclude that 
there are indeed no divergences to be seen in the large $N_T$ limit.  The
normalization in this case is also the expected continuum value.  It is
{\em not} identical to the Stefan-Boltzmann value of 2$\pi^{-2}$.
Using the contour integral method it can be easily shown to be
$\chi^{(4)}_c(0)=2/\pi^2(1+1/4)$, with the additional factor of 0.25 coming
from the term usually cancelled in the usual prescriptions
\cite{hk,kogut,bg,rvg}.   We have found the convergence to the continuum value 
to be strongly $M$ dependent and unfortunately very slow  for all values of $M$,
as seen in the plot B of Fig. (\ref{sus4}).
Introducing the chemical potential by choosing $\hat\mu/s$ as the coefficient of the 
number density term in Eq. (\ref{eqn:wilopmu}), instead of the $\hat\mu/M$
we used, achieves a milder $M$ dependence and a faster convergence towards
the continuum.  Perhaps improving the number density term can achieve a
still faster convergence.

\subsection{A new proposal for QCD critical point via Taylor expansion}

Inspired by the above experience of dealing with the number density in the
linear form, as in Eq. (\ref{eqn:wilopmu}), we make a proposal valid for
all fermions.  Because of the infamous sign/phase problem for the fermion
determinant with nonzero chemical potential, it has been proposed
 to look for the QCD critical point \cite{gg} by looking for the radius
of convergence of the Taylor expansion \cite{bi,gg} in $\mu$ of the baryonic
susceptibility.  Computations have been done up to the eighth 
order so far \cite{gg,gg1}.  Extending these calculations to higher order is both
necessary and desirable to confirm the results already obtained.  Our
proposal can permit such an endeavor.  We denote $M(\mu$) to be  
any generic lattice fermionic operator with the chemical potential $\mu$ :

\begin{eqnarray}
\nonumber 
S_F &=& \sum_{x,y} \bar \Psi (x) M(\mu;x,y) \Psi(y) \\ \label{ref:genD}
    &=& \sum_{x,y} \bar \Psi (x) D(x,y) \Psi(y) 
    + \mu a \sum_{x,y}  N(x,y)
\end{eqnarray}

Here $D$ can be the staggered, overlap, Wilson-Dirac or any other
suitable fermion operator, and $N(x,y)$ is the corresponding point-split and
gauge invariant number density.  Eq. (\ref{eqn:wilopmu}) provides a
concrete example of the above  for the overlap fermions.  Note that any
improvements in the fermion operator $D$ or the number density $N$ are
generically included as long as the classical continuum limit is the same
and $\mu$ appears linearly.

It is easy to see that only the first derivative of $M$ with $\mu$ is
nonzero. All others are zero.  Thus denoting by $M^{'}$ the first derivative 
of $M$ with respect to $\mu$ and adding more primes in the superscript for
successively higher orders, 
\begin{equation}
M' = \sum_{x,y} N(x,y), \qquad {\rm and} \qquad  M'' = M''' = M''''... = 0~,
\end{equation}
for our proposal to incorporate $\mu$ in contrast to the popular 
exp($ \pm \mu$) prescription where {\em all} derivatives are nonzero:
\begin{equation}
\qquad M' = M'''...= \sum_{x,y}N(x,y) \qquad {\rm and} \qquad
M'' = M'''' = M''''''...\ne 0 ~.
\end{equation}

As a consequence, the various nonlinear susceptibility expressions, or
equivalently the expressions for Taylor series coefficients, are a lot
simpler and have a lot fewer terms.  For example, let us write down a
fourth-order coefficient [by combining Eqs. (A.4), (A.7), and (A.8) of
\cite{gg}] :
\begin{equation}
\label{eqn:chi40}
\chi^{(4)}=
\frac{T}{V}\left[ \biggr\langle \mathcal{O}_{1111}+6 \mathcal{O}_{112}+4 \mathcal{O}_{13}+3 \mathcal{O}_{22}+\mathcal{O}_4\biggr\rangle
- 3 \biggr\langle \mathcal{O}_{11}+\mathcal{O}_2\biggr\rangle^2\right]. 
\end{equation} 
Here the notation $\mathcal{O}_{ij\cdots l}$ stands for the product, 
$\mathcal{O}_i\mathcal{O}_j\cdots O_l$.  The expressions for $\mathcal{O}_n$, $n$=1,4 for our proposal
above are 
\begin{eqnarray}
\mathcal{O}_1 &=& \text{Tr~} M^{-1}M', \\ \nonumber
\mathcal{O}_2 &=& - \text{Tr ~} M^{-1}M'M^{-1}M', \\ \nonumber
\mathcal{O}_3 &=& 2 ~\text{Tr ~} (M^{-1}M')^3 , \\ \nonumber
\mathcal{O}_4 &=& - 6 ~\text{Tr ~} (M^{-1}M')^4,
\label{eqn:Ous}
\end{eqnarray}

in contrast with those for the usual case given in \cite {gg} :
\begin{eqnarray}
\mathcal{O}_1 &=& \text{Tr ~} M^{-1}M', \\ \nonumber
\mathcal{O}_2 &=& -\text{Tr ~}  M^{-1}M'M^{-1}M' + \text{Tr ~} M^{-1}M'', \\ \nonumber
\mathcal{O}_3 &=& 2 ~\text{Tr ~} (M^{-1}M')^3  - 3 ~\text{Tr ~} M^{-1}M'M^{-1}M'' + \text{Tr ~} M^{-1}M''', \\ \nonumber
\mathcal{O}_4 &=& - 6~  \text{Tr ~}(M^{-1}M')^4 + 12 ~\text{Tr ~} (M^{-1}M')^2 M^{-1}M'' -3 ~\text{Tr ~}
(M^{-1}M'')^2 \\ \nonumber
&-& 4 ~\text{Tr ~} M^{-1}M'M^{-1}M''' + \text{Tr ~} M^{-1}M''''.
\label{eqn:Ogg}
\end{eqnarray}
The eighth-order term needs $\mathcal{O}_8$, which has 18 terms in the usual case
whereas it will simply be $\mathcal{O}_8 = - 5040 ~ \text{Tr}~(M^{-1}M')^8$ for our proposal.

The number of matrix inversions required to compute the higher order
susceptibilities is also drastically reduced in this way of incorporating the
chemical potential. This would save a considerable amount of computer time, as
matrix inversions are the most time intensive operations. Following 
Fig.  3 of Ref. \cite{gg}, one can see that all computations referred to
on the leftmost branch of the algorithm tree need to be performed when $M$
has a linear $\mu$ dependence. Thus for the eighth-order susceptibility
computation we need to compute only eight matrix inversions as compared to the
20 required there, saving 60\% of the computer time. For higher order
susceptibilities, the number of matrix inversions is reduced by at least
half, enabling us to compute even higher orders of the Taylor series of
thermodynamic quantities and thus constrain the radius of convergence and
the estimated location of the critical point better.

Of course, there is a price to pay, and we hope to demonstrate in the 
future from our ongoing work that it is not very big.  All the coefficients
that one evaluates this way will have the remnants of the terms which are
otherwise eliminated by the usual prescriptions \cite{hk,kogut,bg,rvg}.  Based on
our computations in the previous section, we suggest that the zero
temperature contribution to each of them be subtracted by evaluating
them on a symmetric $N^4$ lattice at the same $\beta = 6/g^2$ as the finite
temperature computation on the $N^3 \times N_T$ lattice.  Since the second-order
 susceptibility $\chi^{(2)}$ has a divergence in the continuum limit,
its computations may need higher precision to ensure the absence of the
cut-off effects but the higher order coefficients have no such
difficulties.  One will also have to rescale the fourth-order
susceptibility by a factor of 1.25 in order to use it in the ratio or the
root method of estimating the radius of convergence.   We hope that
tenth- or even twelfth-order coefficient may thus be computable.

\section{Conclusions}

%The anomaly is an ultraviolet phenomena and it is expected that it should
%not receive any finite temperature and density corrections.  
We have shown
perturbatively from the computation of the triangle diagram at zero
temperature that the anomaly equation does not have any finite density
correction terms.  We have extended our calculations to the nonperturbative
case where we have used Fujikawa's method  to show that the anomaly relation
is unaffected in the presence of a finite chemical potential. This has an
important implication for the lattice field theory in designing the lattice 
Dirac operator for nonzero $\mu$: It should lead to a $\mu$-independent
anomaly relation on the lattice. The recent Bloch-Wettig proposal for chiral fermion
operators at finite density violates the chiral invariance on the lattice 
itself.  While a $\mu$-dependent modification of the chiral transformation
can restore the chiral invariance, it leads to a $\mu$-dependent anomaly
relation unlike in the continuum theory.  Such a modification has other
physical consequences discussed in Ref. \cite{bgsL}.

We have proposed a physically more justified  way of introducing $\mu$ in
the overlap Dirac operator. In this method the chiral symmetry is explicitly
broken as well, but the contribution to the anomaly relation from the measure
 is likely to remain $\mu$ independent,
with the lattice corrections to the anomaly relation falling off as a power law
 in the continuum limit.   It has the expected $\mu^2/a^2$-type divergences in the
continuum limit. We showed how a simple subtraction scheme can take care of
them in the free case.  We proposed to use the simple linear in $\mu$ form
for the Taylor series expansion technique of locating the QCD critical
point.  It has the advantage that the number of fermion matrix inversions
goes down drastically when computing the higher order quark number
susceptibilities. The higher order susceptibility computations are clearly
important to accurately locate the critical point in the $T$-$\mu_B$ phase
space for QCD. Our proposal would save much of the computational effort
required for obtaining higher order susceptibilities, even for the staggered
fermions.

\section*{Acknowledgments}
R.V.G. would like to thank Rajamani Narayanan of FIU, Florida for his
queries about QCD at finite temperature and density which inspired this
work.  S.S. would like to acknowledge the Council of Scientific and
Industrial Research(CSIR) for financial support. We are indebted to
Parthasarathi Mitra of SINP, Kolkata for carefully reading the earlier
version of our manuscript and for his crucial remarks which led us to the
proof in Sec. II.B.

\end{document}